\documentclass[aps,prl,twocolumn,longbibliography,amsfonts,amssymb,amsmath]{revtex4-1}
\usepackage{graphicx,xcolor,soul}

\newcommand{\mean}[1]{\langle #1\rangle}
\newcommand{\unit}[1]{\,\mathrm{#1}}
\renewcommand{\vec}[1]{\mathbf{#1}}
\DeclareMathOperator{\erfc}{erfc}

\newcommand{\al}{\alpha}
\newcommand{\gam}{\gamma}
\newcommand{\kap}{\kappa}
\newcommand{\x}{\vec r}
\newcommand{\nois}{\boldsymbol\eta}
\newcommand{\Def}{D_\text{eff}}
\newcommand{\vs}{v_\text{s}}

\begin{document}

\title{Self-assembly of colloidal molecules due to self-generated flow}

\author{Ran Niu, Thomas Speck, Thomas Palberg}
\affiliation{Institut f\"ur Physik, Johannes Gutenberg-Universit\"at Mainz,
  Staudingerweg 7-9, 55128 Mainz, Germany}

\begin{abstract}
  The emergence of structure through aggregation is a fascinating topic and of both fundamental and practical interest. Here we demonstrate that self-generated solvent flow can be used to generate long-range attractions on the colloidal scale, with sub-pico Newton forces extending into the millimeter-range. We observe a rich dynamic behavior with the formation and fusion of small clusters resembling molecules, the dynamics of which is governed by an effective conservative energy that decays as $1/r$. Breaking the flow symmetry, these clusters can be made active.
\end{abstract}

\maketitle


Colloidal particles acting as ``big'' artificial atoms have been instrumental in studying microscopic processes in condensed matter, from the kinetics of crystallization~\cite{palb14} to the vapor-liquid interface~\cite{aart04}. Due to their size, colloidal particles are observable directly in real space. Moreover, interactions are widely tunable, ranging from hard spheres to long-range repulsive, short-range attractive, and dipolar~\cite{roya03,yeth07}. Consequently, colloidal particles can be assembled into a multitude of different structures: from clusters~\cite{meng10,perr15,demi15} and stable molecules~\cite{mano03,saca10} composed of a few particles to extended bulk structures like ionic binary crystals~\cite{leun05}. In addition, self-assembly into useful superstructures can be controlled by factors such as confinement~\cite{nijs15} and particle shape~\cite{glot07}, which make colloids a versatile and fascinating form of matter~\cite{mano15}.

What is still missing are truly long-range attractions of like-charged (or uncharged) identical colloidal particles. There is much interest in the basic statistical physics of systems with such interactions, which play a role in gravitational collapse, two-dimensional elasticity, chemotactic collapse, quantum fluids, and atomic clusters~\cite{camp09}. One proposed realization are colloidal particles trapped at an interface~\cite{ghez97} that experience screened, long-range attractions due to capillary fluctuations of the interface~\cite{blei11}. The attractive interactions then correspond to Newtonian gravity in two dimensions. Complex patterns are also known to arise for bacteria due to long-range chemotactic interactions~\cite{budr91}. Critical long-range Casimir forces have been reported for colloidal particles in a binary solvent~\cite{hert08}, which are tunable by temperature and surface chemistry. Finally, a recent theoretical proposal are catalytically \emph{active} colloidal particles that interact through producing or consuming chemicals~\cite{soto14,soto15}. For simple diffusion the concentration profile of a chemical decays as inverse distance, implying long-range interactions that can be tuned through activity (how chemicals are produced or consumed) and mobility (how particles react to gradients).

\begin{figure}[b!]
  \centering
  \includegraphics{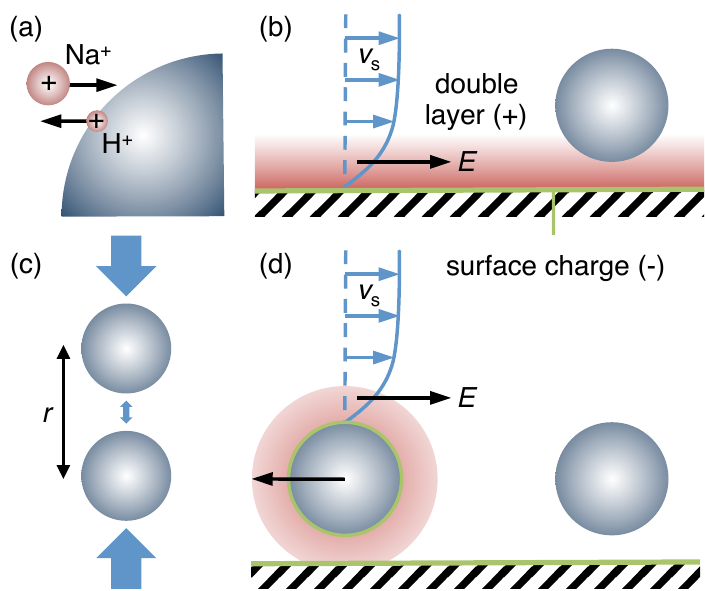}
  \caption{Long-range interactions through flow. (a)~The colloidal particles release hydrogen ions (H$^+$), which are exchanged with residual cationic impurities (here Na$^+$). Different ion mobilities generate local electric fields $E$ that slow the hydrogen ions and accelerate the impurities to maintain overal electro-neutrality. (b)~In the double layer of the substrate these fields generate electro-osmotic flow with solvent velocity $\vs$ towards the particle. (c)~For two particles a distance $r$ apart, the higher H$^+$ concentration in the space between particles reduces the gradient and leads to an asymmetric flow. (d)~In the double layer of the particles, the same effect leads to electro-phoretic flow.}
  \label{fig:mech}
\end{figure}

Here, we implement long-range attractions through hydrodynamic flows coupling suspended particles~\cite{bane16,feld16}. We report on experiments using spherical ion exchange resin particles sedimented to the negatively charged substrate. The particles have diameters of $15\unit{\mu m}$, for which Brownian diffusion is practically negligible on the experimental time scale. They interact due to self-generated local \emph{flow}, resulting in an effective long-range $1/r$ attraction as expected for three-dimensional unscreened gravity. Our present understanding of the mechanism responsible for their aggregation can be summarized as follows~\cite{ande89,niu16}: By exchanging residual cationic impurities for stored hydrogen ions (Fig.~\ref{fig:mech}a), the particles generate a concentration profile $c$ that decays away from the particles. Different diffusion coefficients of the exchanged cations and the released ions locally generate diffusio-electric fields which retain overall electro-neutrality by slowing the outward drift of hydrogen ions and accelerating the inward drift of impurities.


\begin{figure}[b!]
  \centering
  \includegraphics{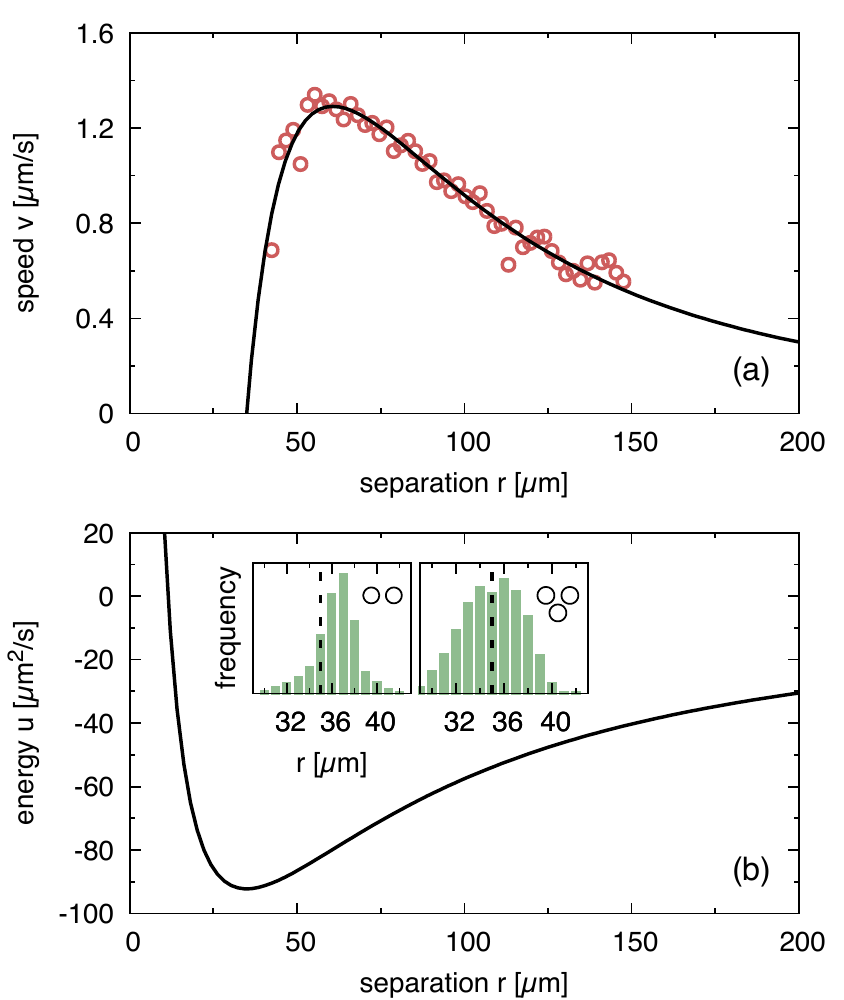}
  \caption{Effective pair potential. (a)~Drift velocity $v(r)$ of two particles with separation $r$ (symbols). The line is the free fit to $v(r)=2u'(r)$ with $u'(r)$ denoting the derivative of Eq.~(\ref{eq:u}). (b)~The resulting effective pair potential $u(r)$. Insets: Histograms of the fluctuating bond length of dimers (left) and trimers (right), where the dashed line indicates the minimum $r_0\simeq35\unit{\mu m}$ of the fitted effective potential.}
  \label{fig:pot}
\end{figure}

\begin{figure*}[t]
  \centering
  \includegraphics{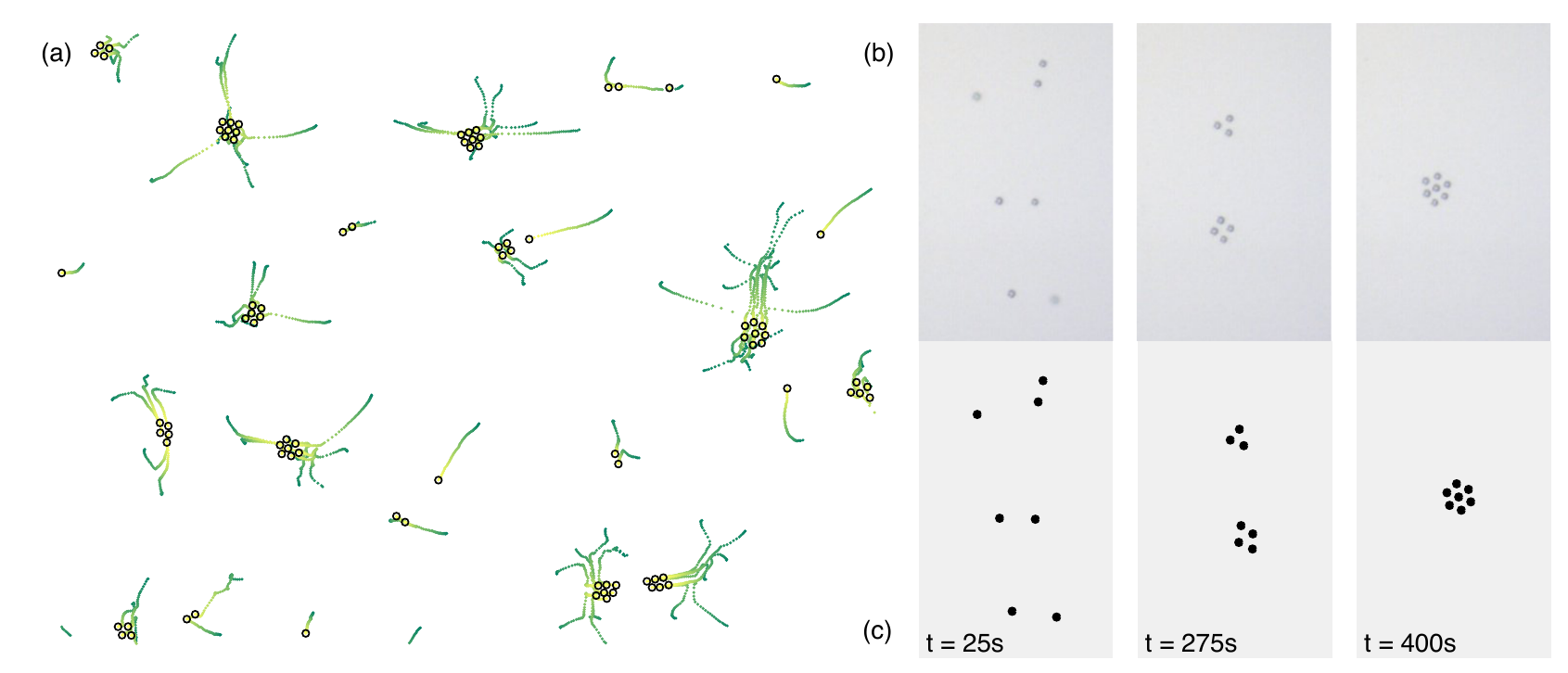}
  \caption{Dynamics of assembly. (a)~Experimental particle traces and the configuration reached after $5\unit{min}$ (discs). (b,c)~Consecutive snapshots from experiment (b) and a single simulation run (c) showing the formation of two clusters in their ground state (middle) and a large cluster with $n=7$ particles (right). The simulations have been initialized with the positions extracted from the first experimental snapshot.}
  \label{fig:dyn}
\end{figure*}

The total system composed of solvent, ions, and colloidal particles is clearly out of thermal equilibrium through the free energy released by solvating the hydrogen ions, which drives the flow. However, appealing to a scale separation between particles and solvent, our crucial assumption will be that the motion of the particles themselves can be described by 
\begin{equation}
  \label{eq:lang}
  \dot\x_i = -\nabla_i U + \nois_i,
\end{equation}
where $U=\sum_{i<j}u(\x_i-\x_j)$ is a conservative potential, $u(r)$ is the pair potential Eq.~(\ref{eq:u}), and $\nois_i$ models the noise with zero mean and correlations $\mean{\nois_i(t)\nois_j^T(t')}=2\Def\mathbf 1\delta_{ij}\delta(t-t')$.

The relative velocity of two particles with separation $\x=\x_1-\x_2$ is $v(r)=\mean{(\x/r)\cdot(\dot\x_1-\dot\x_2)\delta(|\x|-r)}=2u'(r)$ after inserting Eq.~(\ref{eq:lang}). Hence, we have direct access to the interactions $u(r)$ through measuring a dynamic quantity, the approach velocity $v(r)$ as shown in Fig.~\ref{fig:pot}a. Approaching each other, the speed increases as expected but reaches a maximum at about $r\simeq60\unit{\mu m}$ before it drops rapidly. 

This behavior can be rationalized by considering the generated flows in more detail. In the double layer of the substrate, the local fields generate electro-osmotic flow of the solvent since the negative substrate is screened by an excess of positive charges (Fig.~\ref{fig:mech}b). The motion of the particles is determined by the slip velocity $\vs=-\mu_\text{p}\nabla c$ with ion concentration field $c$ and phoretic mobility $\mu_\text{p}$~\cite{ande89}, whereby our measured particle velocities are consistent with a constant mobility. For an isolated particle, the solvent flow is approaching symmetrically and no net motion results (the particle can be regarded as a sink in two dimensions; of course, the solvent is incompressible with backflow out of the plane of the substrate). For two particles, the hydrogen ion concentration is increased in the space between the particles, which reduces the gradient (with respect to the particles' surfaces) and thus the flow velocity. Hence, the flow on each particle now becomes asymmetric, resulting in an apparent attraction (Fig.~\ref{fig:mech}c). Basically the same mechanism acts in the double layer of the particles but now leads to electro-phoretic motion. The particles themselves are (slightly) negatively charged due to the release of cations. The field generated by the concentration gradient of the other particle again generates a solvent flow, however, now particle and solvent taken together are force-free, which leads to a particle speed in the \emph{opposite} direction of the field (Fig.~\ref{fig:mech}d).

For purely diffusive ion motion one would expect the concentration to decay as $c(r)\sim1/r$, but measurements of the pH reveal a more complicated behavior (these measurements had to be performed for a larger particle, see Supplemental Material~\cite{sm}, but we expect the qualitative features to be the same for the smaller particles used here). While there is indeed a $1/r$ decay regime, closer to the ion exchange particle it changes to a slower decay that is well described as exponential. We speculate that this accumulation is caused by the local fields slowing down outward moving ions. We model this effect through a term resembling screening although we stress that it originates from the flow and not electrostatics. Combining both flows, the functional form of the potential reads
\begin{equation}
  \label{eq:u}
  u(r) = -\frac{\gam}{r} + \frac{\al}{r}e^{-r/\xi}
\end{equation}
with three free parameters: the prefactors $\gam$ and $\al$, and the screening length $\xi$. As shown in Fig.~\ref{fig:pot}a, this function describes the experimental data very well. From the fit we obtain $\gam\simeq6120\unit{\mu m^3/s}$, $\al\simeq8805\unit{\mu m^3/s}$, and $\xi\simeq31.4\unit{\mu m}$. The pair potential $u(r)$ plotted in Fig.~\ref{fig:pot}b has a minimum at $r_0\simeq35\unit{\mu m}$. As shown in the inset of Fig.~\ref{fig:pot}b, $r_0$ agrees well with the maximum of the distributions of bond length $r$ for dimers and trimers.

Usually, overdamped motion is described by a product of particle mobility and the gradient of the potential energy. Since we do not have access to these terms separately, we treat $u(r)$ as an effective ``energy'' absorbing the phoretic mobility $\mu$, with $u(r)$ thus having units of a diffusion coefficient. Nevertheless, we can estimate physical energies employing the bare particle mobility $\mu_0$, which quantifies the forces needed to move a single particle through the solvent with desired speed. For our particles in water its value is $\mu_0\simeq7.8\unit{\mu m/(s\cdot pN)}$, yielding forces between particles of order $0.1\unit{pN}$. With $u(r_0)\simeq-92.34\unit{\mu m^2/s}$, the corresponding bond dissociation energy would thus be $E_\text{b}=|u(r_0)|/\mu_0\simeq7000\unit{kJ/mol}$.

The final ingredient for our theoretical model is an effective temperature, which we extract from the measured bond fluctuations. First, from the distribution of the bond length $r$ for dimers we determine its variance $\text{Var}(r)\simeq3.57\unit{\mu m^2}$. Assuming that these vibrations are effectively equilibrated allows us to determine the analog of a temperature, $\Def\approx\text{Var}(r)u''(r_0)\simeq0.3\unit{\mu m^2/s}$. We test this assumption for three particles, for which a quick calculation predicts that the harmonic bond fluctuations are $\frac{5}{3}$ times larger~\cite{sm}. The predicted value $5.95\unit{\mu m^2}$ is only slightly smaller than the measured variance $6.25\unit{\mu m^2}$, the difference of which is due to anharmonic higher-order vibrations.

Our suspension of ion exchange particles is not stationary but slowly collapses to a close-packed state due to the long-range interactions. Starting from a homogeneous density profile, during this process we observe the formation of colloidal clusters (``molecules'') with $n$ particles. This assembly happens autonomously in contrast to prefabricated colloidal molecules~\cite{ebbe10,ni17}. While metastable, single clusters can be observed up to minutes, which allows in principle to study in detail different isomers and the ``reactions'' by which larger clusters form (more details are given as Supplemental Material~\cite{sm}). In Fig.~\ref{fig:dyn}a we show the first few minutes of this process for a dilute suspension. The experiments were carried out with $60-90$ particles within a field of view corresponding to an area fraction of approx. $0.25\%$. Without adjustable parameters, the observed dynamics are reproduced through the model described by Eq.~(\ref{eq:lang}). In Fig.~\ref{fig:dyn}b we show a sequence of experimental snapshots for seven particles. We then perform simulations of the model using the extracted particle positions from the first experimental frame as initial positions. As shown in Fig.~\ref{fig:dyn}c, the simulations agree with the experiments on the same time scale. While for repeated simulation runs the positions differ due to the noise, the average behavior is consistent.

\begin{figure}[t]
  \centering
  \includegraphics{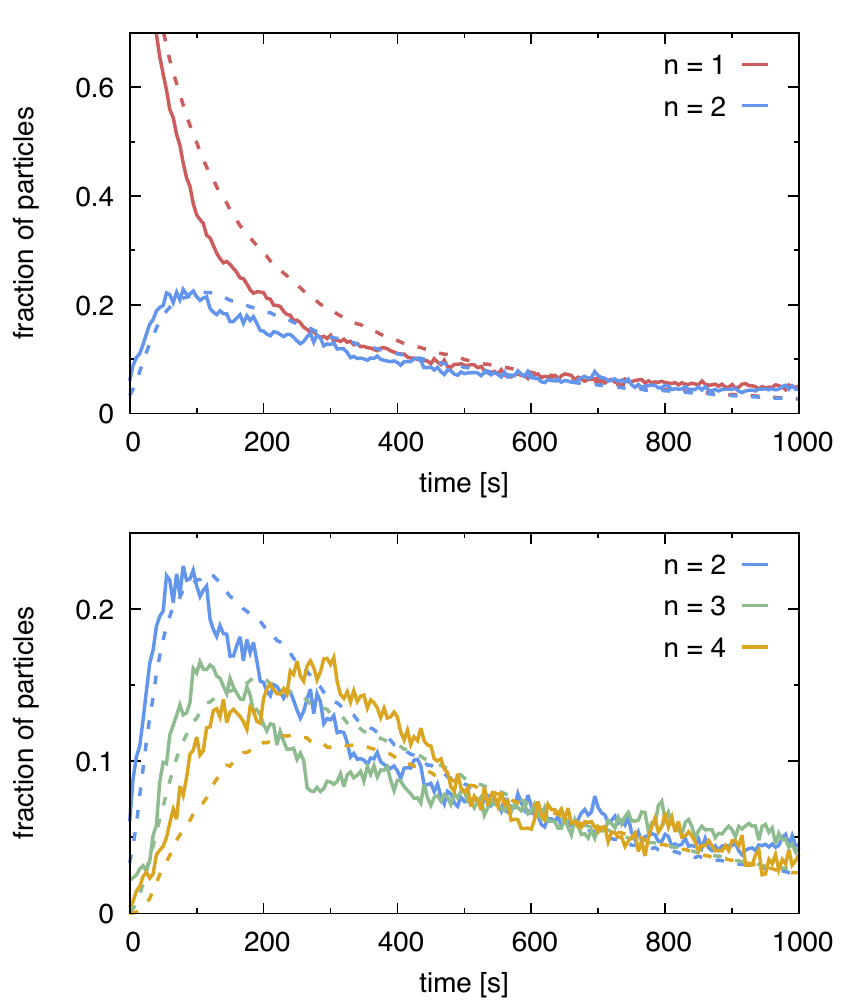}
  \caption{Time evolution of cluster populations $N_n(t)/N$. Solid lines correspond to experimental data (averaged over several experiments), dashed lines to the simulation results of the theoretical model with $N=80$ particles.}
  \label{fig:pop}
\end{figure}

This fit-free quantitative agreement is corroborated by the time evolution of the fraction of clusters with weight $n$. From the analysis of the experiments, we extract the fraction $N_n(t)/N$ of particles residing in clusters with $n$ particles. The range of $N$ is between 60 and 90, and to improve statistics we average over all experiments. We repeat the same analysis with the theoretical model for $N=80$, the comparison of which is shown in Figure~\ref{fig:pop}. The qualitative behavior is that of irreversible aggregation as described through the Smoluchowski coagulation equation~\cite{aldo99}, with a steady decrease of the monomer concentration and peaks for the $n$-mers that become flatter and shifted to later times for increasing weight $n$.

\begin{figure}[t]
  \centering
  \includegraphics{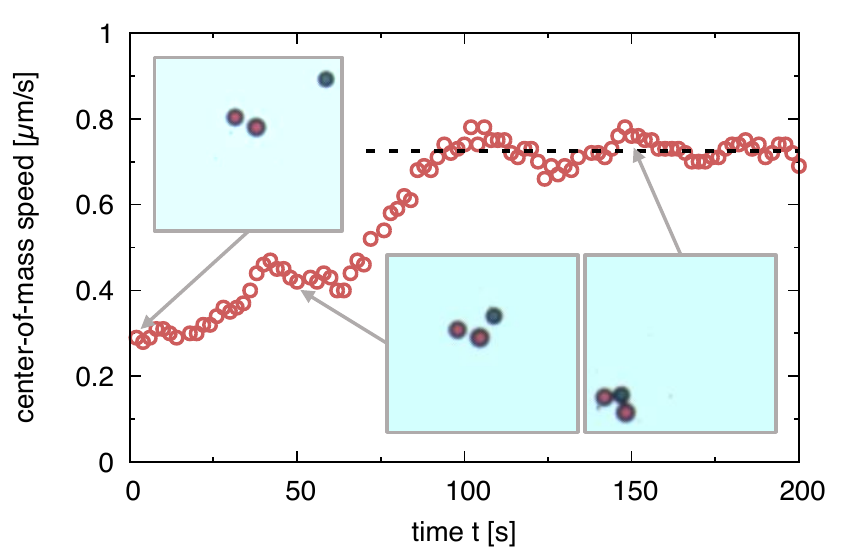}
  \caption{Transition to self-propelled motion. Plotted is the center-of-mass speed of three particles, two cationic and one anionic, which accelerate once assembled into a trimer (at $t\approx50\unit{s}$). For times $t=0\unit{s}$, $50\unit{s}$, and $150\unit{s}$ snapshots are shown with cationic particles in red and anionic particles in blue. The field of view is the same for all snapshots. The dashed line indicates the final speed of the trimer.}
  \label{fig:sym}
\end{figure}

As demonstrated, for a one-component suspension of ion exchange particles in moderate flow the dynamics of the particles alone is effectively described through a conservative potential. However, for larger aggregates also the flow increases, leading to deviations from the predicted behavior (\emph{e.g.} particles are lifted and pushed into the second layer). With even larger flows, particles are observed to spontaneously break the flow symmetry and become self-propelled. Another strategy is to explicitly break symmetry through mixtures of particles with different sizes or shapes, or mixtures of activated and passive particles~\cite{rein13}. Here we explore the consequences of adding anionic ion exchange particles of similar size but releasing OH$^-$ ions, for which we again observe the assembly of low-weight clusters. This is shown exemplary in Fig.~\ref{fig:sym} for two cationic particles and one anionic particle, which assemble into a trimer. As predicted in Ref.~\citenum{soto14}, the geometry together with the different mobilities/activities leads to a self-propelled complex. This can be seen through determining the center-of-mass speed of the three particles, which clearly shows a transition to a constant speed once the trimer has assembled.

To conclude, we have shown that ion exchange particles close to a charged substrate generate flows that lead to effectively conservative, long-range interactions. Autonomous flow-driven assembly through long-range attractions might enable novel non-equilibrium materials and strategies in self-assembly, in particular on intermediate scales for which thermal motion has become negligible but which cannot be manipulated directly. Moreover, through changing composition one may generate asymmetric flow. The resulting directed motion of colloidal particles in combination with volume exclusion leads to fascinating dynamic behavior ranging from clustering~\cite{butt13} and the formation of ``living crystals''~\cite{pala13} to schooling and swarming~\cite{yan16}. Our results demonstrate how one can implement strategies to control and engineer interactions and directed motion on the same footing~\cite{zhan16}, which is a step towards designing active particles that can perform dynamical tasks such as transport of cargo~\cite{sund08,bara12}. Concerning the size of the particles used here, we note that there is no conceptual barrier to using smaller particles, the main issue being the ion exchange rate and ion capacity of singe particles that determines the flow strength and the time over which the flow is being generated.


\acknowledgments

We acknowledge the DFG for funding within the priority program SPP 1726 (grant numbers PA 459/18-1 and SP 1382/3-1).


%

\newpage

\renewcommand{\figurename}{Figure}

\begin{widetext}
  \begin{center}
    {\large\textbf{Supplementary Information}}
  \end{center}
\end{widetext}


\section{Experiments}

The particles are cationic ion exchange resin particles (CK10S, Mitsubishi Chemical Corporation, Japan) with diameter $2a=15.3\pm3.0\unit{\mu m}$ and counter ions Na$^+$. Prior to experiments, the particles were washed with $20\,wt \%$ hydrochloride acid solution to exchange the counter ions into H$^+$. Then we rinsed with doubly deionized water several times until solution pH of $\sim 7$ and dried at $80^\circ\unit{C}$ for 2 h. The sample cell was built from a circular Perspex ring (inner diameter of $20\unit{mm}$, height of $1\unit{mm}$) fixed to microscopy slides by hydrolytically inert epoxy glue and dried for 24 h before use. The glass slides were washed with alkaline solution (Hellmanex\textregistered III, Hellma Analytics) for 30 min under sonication, and subsequently rinsed with tap water and deionized water for several times.

In a typical experiment, a weighted amount of particles was dispersed into doubly deionized water. Subsequently, 400 $\mu$-liters of particle suspension was added into the sample cell. The cell was covered immediately to avoid contamination. Particles settled to the bottom of the cell within minutes. Movies were then taken at a frame rate of 0.5 Hz using an inverted scientific microscope (DMIRBE by Leica, Germany). Particles were tracked through extracting the perimeter using a home-written Python script. The velocity $v(r)$ of two approaching particles was averaged over 160 pairs with any other particle at least 10 times their diameter away.

In a second set of experiments we have added anionic ion exchange resin particles (CA08S, Mitsubishi Chemical Corporation, Japan) with diameter $2a=15.1\pm0.3\unit{\mu m}$ and counter ions Cl$^-$. Before use, particles were washed with concentrated sodium hydroxide (NaOH) to exchange the counter ions into OH$^-$ and subsequently washed with deionized water until the pH of the solution reached about 7. To distinguish cationic ion exchange resin and anionic ion exchange resin, tiny amount of pH indicator solution (pH 4-10, Sigma-Aldrich) was added to mark them redish and bluish, respectively.


\section{Hydrogen ion concentration}

\begin{figure}[t]
  \centering
  \includegraphics{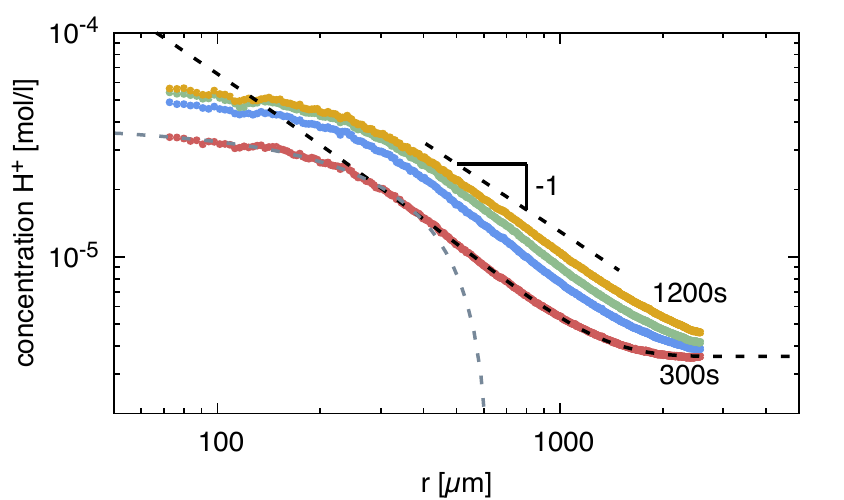}
  \caption{Concentration profiles $c(r,t)=10^{-\text{pH}}$ of hydrogen ions as a function of distance $r$ from the ion exchange particle for different times $t=300,660,900,1200\unit{s}$ (from bottom to top) after contact with water. Indicated are a fit to Eq.~(\ref{eq:c}) (black dashed line), the initial exponential decay (gray dashed line), and the $1/r$ decay (straight dashed line) before the background concentration is approached.}
  \label{fig:pH}
\end{figure}

We determine the pH profile around an ion exchange particle with diameter $45\unit{\mu m}$ using a mixture of Universal indicator solutions (1:3 volume ratio of pH 0-5 and pH 4-10, Sigma-Aldrich, Inc). As the concentration of proton decreases, the color ratio of blue-to-red decreases monotonically. Thus, measuring the blue-to-red color ratio at fixed pH, we get a calibration curve. Applying this calibration curve to every pixel around ion exchange particle, we determine the proton concentration. The color images were recorded using a consumer DSLR (D700, Nikon, Japen) mounted on an inverted scientific microscope (DMIRBE, Leica, Germany). Image recording starts $\sim2\unit{s}$ after the ion exchange particle gets into contact with the indicator solution. The measured concentration $c(r,t)$ as a function of distance $r$ from the particle center and time $t$ after preparation is shown in Figure~\ref{fig:pH}.

A prediction for the concentration profile is given by the solution of the diffusion equation
\begin{equation*}
  \partial_tc(\x,t) = D_+\nabla^2c(\x,t) + \gam\delta(\x)
\end{equation*}
with ion diffusion coefficient $D_+$ and assuming a constant rate $\gam$ with which ions are released. The solution is
\begin{equation}
  \label{eq:c}
  c(r,t) = \frac{\gam}{4\pi D_+r}\erfc\left(\frac{r}{2\sqrt{D_+t}}\right)
  + c_\infty
\end{equation}
with background concentration $c_\infty$. From the fit of this expression to the decay of the concentration profile we obtain a background $\text{pH}\simeq5.45$ and a diffusion coefficient $D_+\simeq1300\unit{\mu m^2/s}$, which is reasonable for hydrogen ions given the fact that they are hydrated. Strikingly, the qualitative behavior changes when approaching the particle and the concentration profile decays much slower than what is expected from the simple diffusion picture. In fact, it can be fitted well by an exponential decay. This implies that outward moving ions are slowed, presumably through local electric fields generated in conjunction with other ions. The resulting solvent flow is thus more complex than $\vs\sim\nabla c\sim1/r^2$, which is captured by the form of our effective pair potential, see the discussion in the main text.


\section{Bond vibrations in the trimer}

While the suspension itself is non-stationary, for the bond vibrations we assume that an effective equilibrium has been reached described by the Boltzmann factor. We expand the energy
\begin{equation*}
  U \approx U_0 + \frac{1}{2}\sum_{ij}^{2n}H_{ij}\delta x_i\delta x_j = U_0 + \frac{1}{2}\sum_{\al=1}^{2n} \kap_\al q_\al^2,
\end{equation*}
where $\vec H$ is the $2n\times2n$ Hessian matrix of second derivatives of the potential energy evaluated at the minimum energy configuration of the cluster with $n$ particles. It has $2n$ eigenvalues $\kap_\al$, of which three are zero corresponding to translation and rotation in two dimensions. The other eigenvalues are the spring constants of the vibrational modes.

For $n=3$, the minimum energy configuration is an equilateral triangle and the Hessian becomes $\vec H=u''(r_0)\vec A$ with constant matrix $\vec A$ (with non-zero eigenvalues $\frac{3}{2},\frac{3}{2},3$). The harmonic fluctuations of any edge can easily be calculated as
\begin{equation*}
  \mean{|\x_i-\x_j-r_0\vec{\hat\x}_{ij}|^2} = \sum_\al \mean{q_\al^2} =
\sum_\al \frac{\Def}{\kap_\al} = \frac{5}{3}\frac {\Def}{u''(r_0)}
\end{equation*}
as used in the main text.


\section{Metastable molecules}

\begin{figure*}[t]
  \centering
  \includegraphics{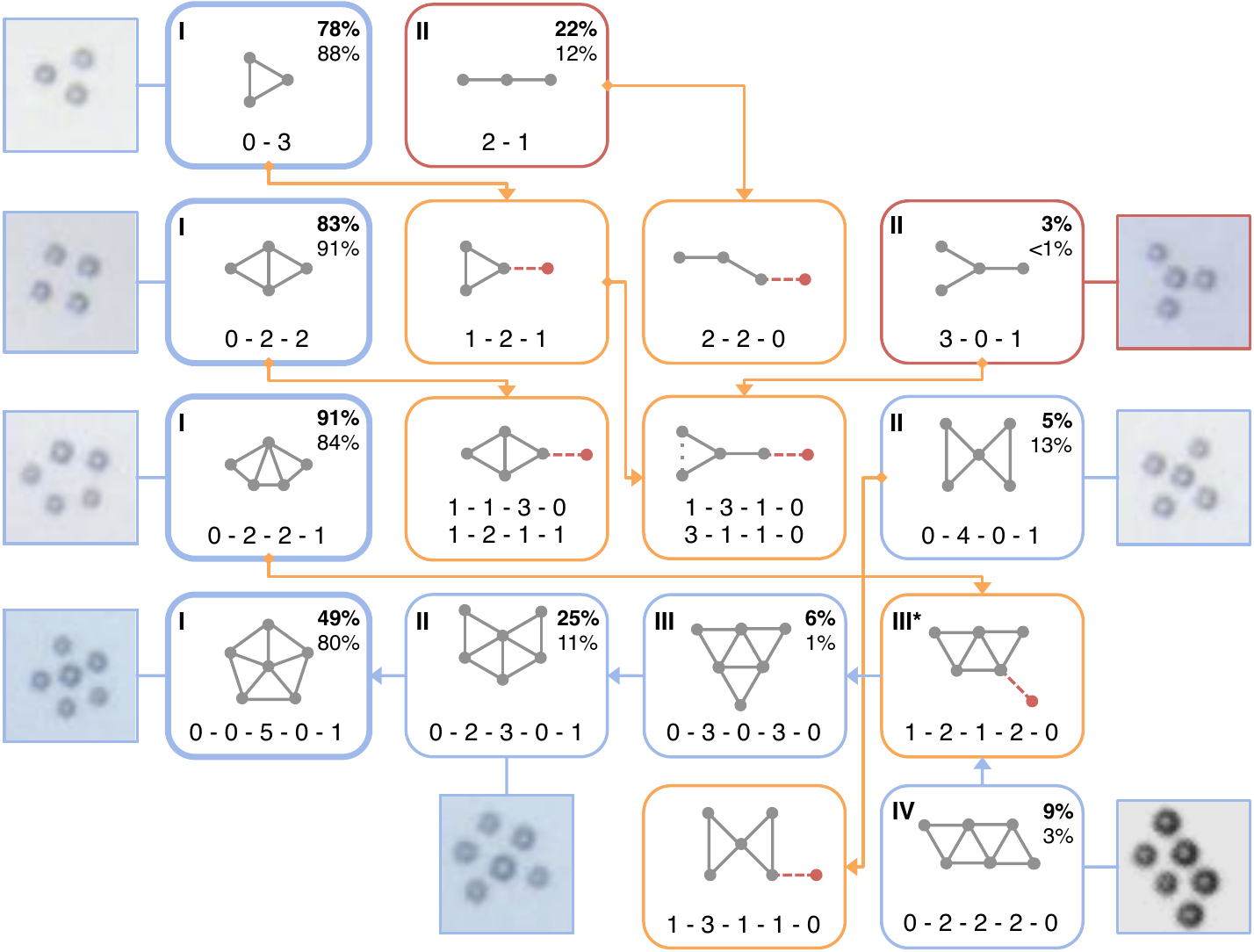}
  \caption{Colloidal ``molecules''. Sketch of molecules with $n$ particles (from top $n=3$ to bottom $n=6$) and snapshots as observed in the experiments. Configurations with blue frames correspond to local energy minima (stable vibrations) whereas red frames indicate unstable configurations. The ground states are indicated to the left with thick frames. Also shown are examples for excited transition states (orange frames), which are characterized by at least one dangling bond (in red). Below every structure the structural fingerprint is given as the vector $n_1$--$n_2$--\dots counting the number of particles with one bond, two bonds, etc. For $n=6$, the blue arrows indicate favored transitions between the stable isomers. In the right corner we compare the population of that structure between experiment (top and bold) and simulation (bottom). Difference to 100\% is the population of transition states.}
  \label{fig:mol}
\end{figure*}

Figure~\ref{fig:mol} shows snapshots of structural arrangements sorted by their molecular weight from $n=3$ to $n=6$. We only show structures that persisted for at least $100\unit{s}$. We distinguish isomers assigning a structural fingerprint counting the number $n_b$ of particles with $b$ direct bonds. For every particle configuration (experiments and simulations) we first determine clusters of mutually bonded particles, where a particle pair $(i,j)$ forms a bond if the separation $|\x_{ij}|$ is smaller than the cutoff of $40\unit{\mu m}$. We then refine the bond network and only retain bonds between \emph{direct} neighbors. A particle $j$ is a direct neighbor of $i$ if for all particles $k$ with distance $|\x_{ik}|<|\x_{ij}|$ the condition $\x_{ik}\cdot\x_{jk}>0$ holds (the enclosed angle is less than $\pi/2$ radians)~\cite{mali13}. For each cluster we count the number of particles $n_b$ with $b$ bonds, the vector of which forms a ``fingerprint'' based on which we identify the isomeric structure of the molecule.

The structures to the left in Figure~\ref{fig:mol} show the ground states minimizing the effective energy. From the eigenspectrum of vibrations we delineate (meta-)stable molecules (blue frames) from unstable molecules (red frames). For $n=3$ and $n=4$ the latter nevertheless occur with a non-vanishing probability since they are populated by reactions adding another monomer. We can identify these transition states easily by their ``dangling bonds'' with $n_1>0$. Transition states quickly rearrange into stable isomers for the same $n$. These are relatively long-lived -- typically until another reaction occurs increasing the molecular weight -- but we also observe spontaneous transitions between isomers (see \emph{Supplementary Movie 2} and Fig.~\ref{fig:trans}). For $n=6$ we have four stable isomers. For short-range attractions counting only direct bonds, the structures II, III, and IV form the degenerate ground state manifold~\cite{perr15}. Due to the long-range attractions in our case, the degeneracy is lifted with the regular pentagon (I) representing the minimal energy configuration. While most transitions proceed through a slight shift of bonds, note that the transition $\text{IV}\to\text{III}$ involves breaking a bond with excited intermediate III$^\ast$. Consequently, for the parallelogram (IV) we observe a higher population both in experiment and theory than one would expect from its energy alone.

\begin{figure}[h]
  \centering
  \includegraphics[width=\linewidth]{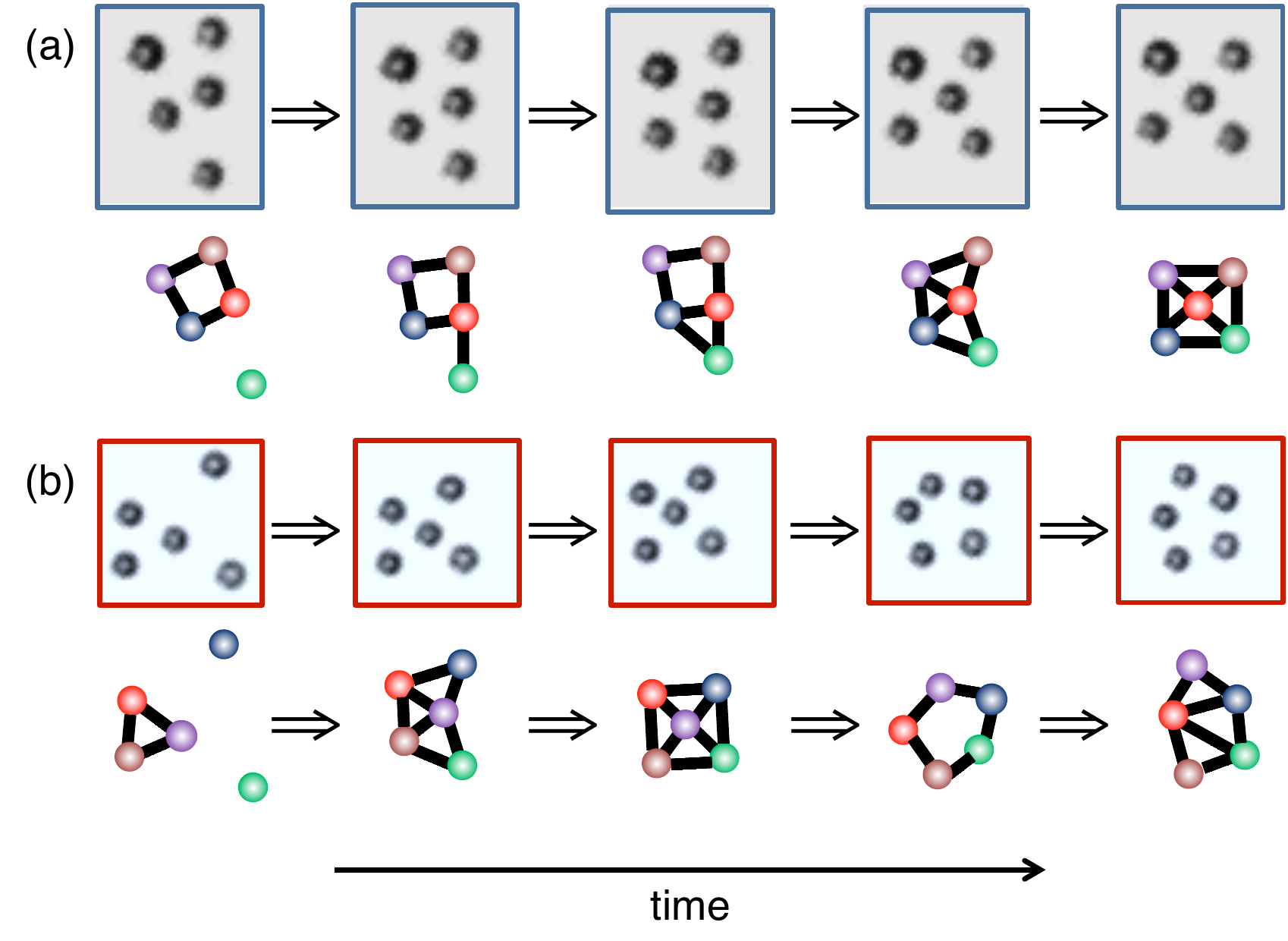}
  \caption{Two examples of structural changes for $n=5$ (see also \emph{Supplementary Movie 2}). (a)~Formation of a transition structure after addition of one monomer, which quickly relaxes to configuration 5-II. (b)~Trimer plus two monomers again relaxing to 5-II but now followed by the reorganization into the ground state 5-I through pushing out the center particle.}
  \label{fig:trans}
\end{figure}

\begin{thebibliography}{36}%
\makeatletter
\providecommand \@ifxundefined [1]{%
 \@ifx{#1\undefined}
}%
\providecommand \@ifnum [1]{%
 \ifnum #1\expandafter \@firstoftwo
 \else \expandafter \@secondoftwo
 \fi
}%
\providecommand \@ifx [1]{%
 \ifx #1\expandafter \@firstoftwo
 \else \expandafter \@secondoftwo
 \fi
}%
\providecommand \natexlab [1]{#1}%
\providecommand \enquote  [1]{``#1''}%
\providecommand \bibnamefont  [1]{#1}%
\providecommand \bibfnamefont [1]{#1}%
\providecommand \citenamefont [1]{#1}%
\providecommand \href@noop [0]{\@secondoftwo}%
\providecommand \href [0]{\begingroup \@sanitize@url \@href}%
\providecommand \@href[1]{\@@startlink{#1}\@@href}%
\providecommand \@@href[1]{\endgroup#1\@@endlink}%
\providecommand \@sanitize@url [0]{\catcode `\\12\catcode `\$12\catcode
  `\&12\catcode `\#12\catcode `\^12\catcode `\_12\catcode `\%12\relax}%
\providecommand \@@startlink[1]{}%
\providecommand \@@endlink[0]{}%
\providecommand \url  [0]{\begingroup\@sanitize@url \@url }%
\providecommand \@url [1]{\endgroup\@href {#1}{\urlprefix }}%
\providecommand \urlprefix  [0]{URL }%
\providecommand \Eprint [0]{\href }%
\providecommand \doibase [0]{http://dx.doi.org/}%
\providecommand \selectlanguage [0]{\@gobble}%
\providecommand \bibinfo  [0]{\@secondoftwo}%
\providecommand \bibfield  [0]{\@secondoftwo}%
\providecommand \translation [1]{[#1]}%
\providecommand \BibitemOpen [0]{}%
\providecommand \bibitemStop [0]{}%
\providecommand \bibitemNoStop [0]{.\EOS\space}%
\providecommand \EOS [0]{\spacefactor3000\relax}%
\providecommand \BibitemShut  [1]{\csname bibitem#1\endcsname}%
\let\auto@bib@innerbib\@empty
\bibitem [{\citenamefont {Palberg}(2014)}]{palb14}%
  \BibitemOpen
  \bibfield  {author} {\bibinfo {author} {\bibfnamefont {T.}~\bibnamefont
  {Palberg}},\ }\bibfield  {title} {\enquote {\bibinfo {title} {Crystallization
  kinetics of colloidal model suspensions: recent achievements and new
  perspectives},}\ }\href {\doibase 10.1088/0953-8984/26/33/333101} {\bibfield
  {journal} {\bibinfo  {journal} {J. Phys.: Condens. Matter}\ }\textbf
  {\bibinfo {volume} {26}},\ \bibinfo {pages} {333101} (\bibinfo {year}
  {2014})}\BibitemShut {NoStop}%
\bibitem [{\citenamefont {Aarts}\ \emph {et~al.}(2004)\citenamefont {Aarts},
  \citenamefont {Schmidt},\ and\ \citenamefont {Lekkerkerker}}]{aart04}%
  \BibitemOpen
  \bibfield  {author} {\bibinfo {author} {\bibfnamefont {D.~G. A.~L.}\
  \bibnamefont {Aarts}}, \bibinfo {author} {\bibfnamefont {M.}~\bibnamefont
  {Schmidt}}, \ and\ \bibinfo {author} {\bibfnamefont {H.~N.~W.}\ \bibnamefont
  {Lekkerkerker}},\ }\bibfield  {title} {\enquote {\bibinfo {title} {Direct
  visual observation of thermal capillary waves},}\ }\href {\doibase
  10.1126/science.1097116} {\bibfield  {journal} {\bibinfo  {journal}
  {Science}\ }\textbf {\bibinfo {volume} {304}},\ \bibinfo {pages} {847--850}
  (\bibinfo {year} {2004})}\BibitemShut {NoStop}%
\bibitem [{\citenamefont {Royall}\ \emph {et~al.}(2003)\citenamefont {Royall},
  \citenamefont {Leunissen},\ and\ \citenamefont {van Blaaderen}}]{roya03}%
  \BibitemOpen
  \bibfield  {author} {\bibinfo {author} {\bibfnamefont {C.~P.}\ \bibnamefont
  {Royall}}, \bibinfo {author} {\bibfnamefont {M.~E.}\ \bibnamefont
  {Leunissen}}, \ and\ \bibinfo {author} {\bibfnamefont {A.}~\bibnamefont {van
  Blaaderen}},\ }\bibfield  {title} {\enquote {\bibinfo {title} {A new
  colloidal model system to study long-range interactions quantitatively in
  real space},}\ }\href@noop {} {\bibfield  {journal} {\bibinfo  {journal} {J.
  Phys. Condens. Matter}\ }\textbf {\bibinfo {volume} {15}},\ \bibinfo {pages}
  {S3581} (\bibinfo {year} {2003})}\BibitemShut {NoStop}%
\bibitem [{\citenamefont {Yethiraj}(2007)}]{yeth07}%
  \BibitemOpen
  \bibfield  {author} {\bibinfo {author} {\bibfnamefont {A.}~\bibnamefont
  {Yethiraj}},\ }\bibfield  {title} {\enquote {\bibinfo {title} {Tunable
  colloids: control of colloidal phase transitions with tunable
  interactions},}\ }\href {\doibase 10.1039/B704251P} {\bibfield  {journal}
  {\bibinfo  {journal} {Soft Matter}\ }\textbf {\bibinfo {volume} {3}},\
  \bibinfo {pages} {1099--1115} (\bibinfo {year} {2007})}\BibitemShut {NoStop}%
\bibitem [{\citenamefont {Meng}\ \emph {et~al.}(2010)\citenamefont {Meng},
  \citenamefont {Arkus}, \citenamefont {Brenner},\ and\ \citenamefont
  {Manoharan}}]{meng10}%
  \BibitemOpen
  \bibfield  {author} {\bibinfo {author} {\bibfnamefont {G.}~\bibnamefont
  {Meng}}, \bibinfo {author} {\bibfnamefont {N.}~\bibnamefont {Arkus}},
  \bibinfo {author} {\bibfnamefont {M.~P.}\ \bibnamefont {Brenner}}, \ and\
  \bibinfo {author} {\bibfnamefont {V.~N.}\ \bibnamefont {Manoharan}},\
  }\bibfield  {title} {\enquote {\bibinfo {title} {The free-energy landscape of
  clusters of attractive hard spheres},}\ }\href {\doibase
  10.1126/science.1181263} {\bibfield  {journal} {\bibinfo  {journal}
  {Science}\ }\textbf {\bibinfo {volume} {327}},\ \bibinfo {pages} {560--563}
  (\bibinfo {year} {2010})}\BibitemShut {NoStop}%
\bibitem [{\citenamefont {Perry}\ \emph {et~al.}(2015)\citenamefont {Perry},
  \citenamefont {Holmes-Cerfon}, \citenamefont {Brenner},\ and\ \citenamefont
  {Manoharan}}]{perr15}%
  \BibitemOpen
  \bibfield  {author} {\bibinfo {author} {\bibfnamefont {R.~W.}\ \bibnamefont
  {Perry}}, \bibinfo {author} {\bibfnamefont {M.~C.}\ \bibnamefont
  {Holmes-Cerfon}}, \bibinfo {author} {\bibfnamefont {M.~P.}\ \bibnamefont
  {Brenner}}, \ and\ \bibinfo {author} {\bibfnamefont {V.~N.}\ \bibnamefont
  {Manoharan}},\ }\bibfield  {title} {\enquote {\bibinfo {title}
  {Two-dimensional clusters of colloidal spheres: Ground states, excited
  states, and structural rearrangements},}\ }\href {\doibase
  10.1103/PhysRevLett.114.228301} {\bibfield  {journal} {\bibinfo  {journal}
  {Phys. Rev. Lett.}\ }\textbf {\bibinfo {volume} {114}},\ \bibinfo {pages}
  {228301} (\bibinfo {year} {2015})}\BibitemShut {NoStop}%
\bibitem [{\citenamefont {Demir\"ors}\ \emph {et~al.}(2015)\citenamefont
  {Demir\"ors}, \citenamefont {Stiefelhagen}, \citenamefont {Vissers},
  \citenamefont {Smallenburg}, \citenamefont {Dijkstra}, \citenamefont
  {Imhof},\ and\ \citenamefont {van Blaaderen}}]{demi15}%
  \BibitemOpen
  \bibfield  {author} {\bibinfo {author} {\bibfnamefont {A.~F.}\ \bibnamefont
  {Demir\"ors}}, \bibinfo {author} {\bibfnamefont {J.~C.~P.}\ \bibnamefont
  {Stiefelhagen}}, \bibinfo {author} {\bibfnamefont {T.}~\bibnamefont
  {Vissers}}, \bibinfo {author} {\bibfnamefont {F.}~\bibnamefont
  {Smallenburg}}, \bibinfo {author} {\bibfnamefont {M.}~\bibnamefont
  {Dijkstra}}, \bibinfo {author} {\bibfnamefont {A.}~\bibnamefont {Imhof}}, \
  and\ \bibinfo {author} {\bibfnamefont {A.}~\bibnamefont {van Blaaderen}},\
  }\bibfield  {title} {\enquote {\bibinfo {title} {Long-ranged oppositely
  charged interactions for designing new types of colloidal clusters},}\ }\href
  {\doibase 10.1103/PhysRevX.5.021012} {\bibfield  {journal} {\bibinfo
  {journal} {Phys. Rev. X}\ }\textbf {\bibinfo {volume} {5}},\ \bibinfo {pages}
  {021012} (\bibinfo {year} {2015})}\BibitemShut {NoStop}%
\bibitem [{\citenamefont {Manoharan}\ \emph {et~al.}(2003)\citenamefont
  {Manoharan}, \citenamefont {Elsesser},\ and\ \citenamefont {Pine}}]{mano03}%
  \BibitemOpen
  \bibfield  {author} {\bibinfo {author} {\bibfnamefont {V.~N.}\ \bibnamefont
  {Manoharan}}, \bibinfo {author} {\bibfnamefont {M.~T.}\ \bibnamefont
  {Elsesser}}, \ and\ \bibinfo {author} {\bibfnamefont {D.~J.}\ \bibnamefont
  {Pine}},\ }\bibfield  {title} {\enquote {\bibinfo {title} {Dense packing and
  symmetry in small clusters of microspheres},}\ }\href {\doibase
  10.1126/science.1086189} {\bibfield  {journal} {\bibinfo  {journal}
  {Science}\ }\textbf {\bibinfo {volume} {301}},\ \bibinfo {pages} {483--487}
  (\bibinfo {year} {2003})}\BibitemShut {NoStop}%
\bibitem [{\citenamefont {Sacanna}\ \emph {et~al.}(2010)\citenamefont
  {Sacanna}, \citenamefont {Irvine}, \citenamefont {Chaikin},\ and\
  \citenamefont {Pine}}]{saca10}%
  \BibitemOpen
  \bibfield  {author} {\bibinfo {author} {\bibfnamefont {S.}~\bibnamefont
  {Sacanna}}, \bibinfo {author} {\bibfnamefont {W.~T.~M.}\ \bibnamefont
  {Irvine}}, \bibinfo {author} {\bibfnamefont {P.~M.}\ \bibnamefont {Chaikin}},
  \ and\ \bibinfo {author} {\bibfnamefont {D.~J.}\ \bibnamefont {Pine}},\
  }\bibfield  {title} {\enquote {\bibinfo {title} {Lock and key colloids},}\
  }\href {\doibase 10.1038/nature08906} {\bibfield  {journal} {\bibinfo
  {journal} {Nature}\ }\textbf {\bibinfo {volume} {464}},\ \bibinfo {pages}
  {575--578} (\bibinfo {year} {2010})}\BibitemShut {NoStop}%
\bibitem [{\citenamefont {Leunissen}\ \emph {et~al.}(2005)\citenamefont
  {Leunissen}, \citenamefont {Christova}, \citenamefont {Hynninen},
  \citenamefont {Royall}, \citenamefont {Campbell}, \citenamefont {Imhof},
  \citenamefont {Dijkstra}, \citenamefont {van Roij},\ and\ \citenamefont {van
  Blaaderen}}]{leun05}%
  \BibitemOpen
  \bibfield  {author} {\bibinfo {author} {\bibfnamefont {M.~E.}\ \bibnamefont
  {Leunissen}}, \bibinfo {author} {\bibfnamefont {C.~G.}\ \bibnamefont
  {Christova}}, \bibinfo {author} {\bibfnamefont {A.-P.}\ \bibnamefont
  {Hynninen}}, \bibinfo {author} {\bibfnamefont {C.~P.}\ \bibnamefont
  {Royall}}, \bibinfo {author} {\bibfnamefont {A.~I.}\ \bibnamefont
  {Campbell}}, \bibinfo {author} {\bibfnamefont {A.}~\bibnamefont {Imhof}},
  \bibinfo {author} {\bibfnamefont {M.}~\bibnamefont {Dijkstra}}, \bibinfo
  {author} {\bibfnamefont {R.}~\bibnamefont {van Roij}}, \ and\ \bibinfo
  {author} {\bibfnamefont {A.}~\bibnamefont {van Blaaderen}},\ }\bibfield
  {title} {\enquote {\bibinfo {title} {Ionic colloidal crystals of oppositely
  charged particles},}\ }\href {\doibase 10.1038/nature03946} {\bibfield
  {journal} {\bibinfo  {journal} {Nature}\ }\textbf {\bibinfo {volume} {437}},\
  \bibinfo {pages} {235--240} (\bibinfo {year} {2005})}\BibitemShut {NoStop}%
\bibitem [{\citenamefont {de~Nijs}\ \emph {et~al.}(2015)\citenamefont
  {de~Nijs}, \citenamefont {Dussi}, \citenamefont {Smallenburg}, \citenamefont
  {Meeldijk}, \citenamefont {Groenendijk}, \citenamefont {Filion},
  \citenamefont {Imhof}, \citenamefont {van Blaaderen},\ and\ \citenamefont
  {Dijkstra}}]{nijs15}%
  \BibitemOpen
  \bibfield  {author} {\bibinfo {author} {\bibfnamefont {B.}~\bibnamefont
  {de~Nijs}}, \bibinfo {author} {\bibfnamefont {S.}~\bibnamefont {Dussi}},
  \bibinfo {author} {\bibfnamefont {F.}~\bibnamefont {Smallenburg}}, \bibinfo
  {author} {\bibfnamefont {J.~D.}\ \bibnamefont {Meeldijk}}, \bibinfo {author}
  {\bibfnamefont {D.~J.}\ \bibnamefont {Groenendijk}}, \bibinfo {author}
  {\bibfnamefont {L.}~\bibnamefont {Filion}}, \bibinfo {author} {\bibfnamefont
  {A.}~\bibnamefont {Imhof}}, \bibinfo {author} {\bibfnamefont
  {A.}~\bibnamefont {van Blaaderen}}, \ and\ \bibinfo {author} {\bibfnamefont
  {M.}~\bibnamefont {Dijkstra}},\ }\bibfield  {title} {\enquote {\bibinfo
  {title} {Entropy-driven formation of large icosahedral colloidal clusters by
  spherical confinement},}\ }\href {\doibase 10.1038/nmat4072} {\bibfield
  {journal} {\bibinfo  {journal} {Nature Mater.}\ }\textbf {\bibinfo {volume}
  {14}},\ \bibinfo {pages} {56--60} (\bibinfo {year} {2015})}\BibitemShut
  {NoStop}%
\bibitem [{\citenamefont {Glotzer}\ and\ \citenamefont
  {Solomon}(2007)}]{glot07}%
  \BibitemOpen
  \bibfield  {author} {\bibinfo {author} {\bibfnamefont {S.~C.}\ \bibnamefont
  {Glotzer}}\ and\ \bibinfo {author} {\bibfnamefont {M.~J.}\ \bibnamefont
  {Solomon}},\ }\bibfield  {title} {\enquote {\bibinfo {title} {Anisotropy of
  building blocks and their assembly into complex structures},}\ }\href
  {\doibase 10.1038/nmat1949} {\bibfield  {journal} {\bibinfo  {journal}
  {Nature Mater.}\ }\textbf {\bibinfo {volume} {6}},\ \bibinfo {pages}
  {557--562} (\bibinfo {year} {2007})}\BibitemShut {NoStop}%
\bibitem [{\citenamefont {Manoharan}(2015)}]{mano15}%
  \BibitemOpen
  \bibfield  {author} {\bibinfo {author} {\bibfnamefont {V.~N.}\ \bibnamefont
  {Manoharan}},\ }\bibfield  {title} {\enquote {\bibinfo {title} {Colloidal
  matter: Packing, geometry, and entropy},}\ }\href {\doibase
  10.1126/science.1253751} {\bibfield  {journal} {\bibinfo  {journal}
  {Science}\ }\textbf {\bibinfo {volume} {349}} (\bibinfo {year} {2015}),\
  10.1126/science.1253751}\BibitemShut {NoStop}%
\bibitem [{\citenamefont {Campa}\ \emph {et~al.}(2009)\citenamefont {Campa},
  \citenamefont {Dauxois},\ and\ \citenamefont {Ruffo}}]{camp09}%
  \BibitemOpen
  \bibfield  {author} {\bibinfo {author} {\bibfnamefont {A.}~\bibnamefont
  {Campa}}, \bibinfo {author} {\bibfnamefont {T.}~\bibnamefont {Dauxois}}, \
  and\ \bibinfo {author} {\bibfnamefont {S.}~\bibnamefont {Ruffo}},\ }\bibfield
   {title} {\enquote {\bibinfo {title} {Statistical mechanics and dynamics of
  solvable models with long-range interactions},}\ }\href {\doibase
  10.1016/j.physrep.2009.07.001} {\bibfield  {journal} {\bibinfo  {journal}
  {Phys. Rep.}\ }\textbf {\bibinfo {volume} {480}},\ \bibinfo {pages} {57--159}
  (\bibinfo {year} {2009})}\BibitemShut {NoStop}%
\bibitem [{\citenamefont {Ghezzi}\ and\ \citenamefont
  {Earnshaw}(1997)}]{ghez97}%
  \BibitemOpen
  \bibfield  {author} {\bibinfo {author} {\bibfnamefont {F.}~\bibnamefont
  {Ghezzi}}\ and\ \bibinfo {author} {\bibfnamefont {J.~C.}\ \bibnamefont
  {Earnshaw}},\ }\bibfield  {title} {\enquote {\bibinfo {title} {Formation of
  meso-structures in colloidal monolayers},}\ }\href@noop {} {\bibfield
  {journal} {\bibinfo  {journal} {J. Phys. Condens. Matter}\ }\textbf {\bibinfo
  {volume} {9}},\ \bibinfo {pages} {L517} (\bibinfo {year} {1997})}\BibitemShut
  {NoStop}%
\bibitem [{\citenamefont {Bleibel}\ \emph {et~al.}(2011)\citenamefont
  {Bleibel}, \citenamefont {Dietrich}, \citenamefont {Dom\'{\i}nguez},\ and\
  \citenamefont {Oettel}}]{blei11}%
  \BibitemOpen
  \bibfield  {author} {\bibinfo {author} {\bibfnamefont {J.}~\bibnamefont
  {Bleibel}}, \bibinfo {author} {\bibfnamefont {S.}~\bibnamefont {Dietrich}},
  \bibinfo {author} {\bibfnamefont {A.}~\bibnamefont {Dom\'{\i}nguez}}, \ and\
  \bibinfo {author} {\bibfnamefont {M.}~\bibnamefont {Oettel}},\ }\bibfield
  {title} {\enquote {\bibinfo {title} {Shock waves in capillary collapse of
  colloids: A model system for two-dimensional screened newtonian gravity},}\
  }\href {\doibase 10.1103/PhysRevLett.107.128302} {\bibfield  {journal}
  {\bibinfo  {journal} {Phys. Rev. Lett.}\ }\textbf {\bibinfo {volume} {107}},\
  \bibinfo {pages} {128302} (\bibinfo {year} {2011})}\BibitemShut {NoStop}%
\bibitem [{\citenamefont {Budrene}\ and\ \citenamefont {Berg}(1991)}]{budr91}%
  \BibitemOpen
  \bibfield  {author} {\bibinfo {author} {\bibfnamefont {E.~O.}\ \bibnamefont
  {Budrene}}\ and\ \bibinfo {author} {\bibfnamefont {H.~C.}\ \bibnamefont
  {Berg}},\ }\bibfield  {title} {\enquote {\bibinfo {title} {Complex patterns
  formed by motile cells of escherichia coli},}\ }\href {\doibase
  10.1038/349630a0} {\bibfield  {journal} {\bibinfo  {journal} {Nature}\
  }\textbf {\bibinfo {volume} {349}},\ \bibinfo {pages} {630--633} (\bibinfo
  {year} {1991})}\BibitemShut {NoStop}%
\bibitem [{\citenamefont {Hertlein}\ \emph {et~al.}(2008)\citenamefont
  {Hertlein}, \citenamefont {Helden}, \citenamefont {Gambassi}, \citenamefont
  {Dietrich},\ and\ \citenamefont {Bechinger}}]{hert08}%
  \BibitemOpen
  \bibfield  {author} {\bibinfo {author} {\bibfnamefont {C.}~\bibnamefont
  {Hertlein}}, \bibinfo {author} {\bibfnamefont {L.}~\bibnamefont {Helden}},
  \bibinfo {author} {\bibfnamefont {A.}~\bibnamefont {Gambassi}}, \bibinfo
  {author} {\bibfnamefont {S.}~\bibnamefont {Dietrich}}, \ and\ \bibinfo
  {author} {\bibfnamefont {C.}~\bibnamefont {Bechinger}},\ }\bibfield  {title}
  {\enquote {\bibinfo {title} {Direct measurement of critical casimir
  forces},}\ }\href {\doibase 10.1038/nature06443} {\bibfield  {journal}
  {\bibinfo  {journal} {Nature}\ }\textbf {\bibinfo {volume} {451}},\ \bibinfo
  {pages} {172--175} (\bibinfo {year} {2008})}\BibitemShut {NoStop}%
\bibitem [{\citenamefont {Soto}\ and\ \citenamefont
  {Golestanian}(2014)}]{soto14}%
  \BibitemOpen
  \bibfield  {author} {\bibinfo {author} {\bibfnamefont {R.}~\bibnamefont
  {Soto}}\ and\ \bibinfo {author} {\bibfnamefont {R.}~\bibnamefont
  {Golestanian}},\ }\bibfield  {title} {\enquote {\bibinfo {title}
  {Self-assembly of catalytically active colloidal molecules: Tailoring
  activity through surface chemistry},}\ }\href {\doibase
  10.1103/PhysRevLett.112.068301} {\bibfield  {journal} {\bibinfo  {journal}
  {Phys. Rev. Lett.}\ }\textbf {\bibinfo {volume} {112}},\ \bibinfo {pages}
  {068301} (\bibinfo {year} {2014})}\BibitemShut {NoStop}%
\bibitem [{\citenamefont {Soto}\ and\ \citenamefont
  {Golestanian}(2015)}]{soto15}%
  \BibitemOpen
  \bibfield  {author} {\bibinfo {author} {\bibfnamefont {R.}~\bibnamefont
  {Soto}}\ and\ \bibinfo {author} {\bibfnamefont {R.}~\bibnamefont
  {Golestanian}},\ }\bibfield  {title} {\enquote {\bibinfo {title}
  {Self-assembly of active colloidal molecules with dynamic function},}\ }\href
  {\doibase 10.1103/PhysRevE.91.052304} {\bibfield  {journal} {\bibinfo
  {journal} {Phys. Rev. E}\ }\textbf {\bibinfo {volume} {91}},\ \bibinfo
  {pages} {052304} (\bibinfo {year} {2015})}\BibitemShut {NoStop}%
\bibitem [{\citenamefont {Banerjee}\ \emph {et~al.}(2016)\citenamefont
  {Banerjee}, \citenamefont {Williams}, \citenamefont {Azevedo}, \citenamefont
  {Helgeson},\ and\ \citenamefont {Squires}}]{bane16}%
  \BibitemOpen
  \bibfield  {author} {\bibinfo {author} {\bibfnamefont {A.}~\bibnamefont
  {Banerjee}}, \bibinfo {author} {\bibfnamefont {I.}~\bibnamefont {Williams}},
  \bibinfo {author} {\bibfnamefont {R.~N.}\ \bibnamefont {Azevedo}}, \bibinfo
  {author} {\bibfnamefont {M.~E.}\ \bibnamefont {Helgeson}}, \ and\ \bibinfo
  {author} {\bibfnamefont {T.~M.}\ \bibnamefont {Squires}},\ }\bibfield
  {title} {\enquote {\bibinfo {title} {Soluto-inertial phenomena: Designing
  long-range, long-lasting, surface-specific interactions in suspensions},}\
  }\href {\doibase 10.1073/pnas.1604743113} {\bibfield  {journal} {\bibinfo
  {journal} {Proc. Natl. Acad. Sci. U.S.A.}\ }\textbf {\bibinfo {volume}
  {113}},\ \bibinfo {pages} {8612--8617} (\bibinfo {year} {2016})}\BibitemShut
  {NoStop}%
\bibitem [{\citenamefont {Feldmann}\ \emph {et~al.}(2016)\citenamefont
  {Feldmann}, \citenamefont {Maduar}, \citenamefont {Santer}, \citenamefont
  {Lomadze}, \citenamefont {Vinogradova},\ and\ \citenamefont
  {Santer}}]{feld16}%
  \BibitemOpen
  \bibfield  {author} {\bibinfo {author} {\bibfnamefont {D.}~\bibnamefont
  {Feldmann}}, \bibinfo {author} {\bibfnamefont {S.~R.}\ \bibnamefont
  {Maduar}}, \bibinfo {author} {\bibfnamefont {M.}~\bibnamefont {Santer}},
  \bibinfo {author} {\bibfnamefont {N.}~\bibnamefont {Lomadze}}, \bibinfo
  {author} {\bibfnamefont {O.~I.}\ \bibnamefont {Vinogradova}}, \ and\ \bibinfo
  {author} {\bibfnamefont {S.}~\bibnamefont {Santer}},\ }\bibfield  {title}
  {\enquote {\bibinfo {title} {Manipulation of small particles at solid liquid
  interface: light driven diffusioosmosis},}\ }\href {\doibase
  10.1038/srep36443} {\bibfield  {journal} {\bibinfo  {journal} {Sci. Rep.}\
  }\textbf {\bibinfo {volume} {6}},\ \bibinfo {pages} {36443} (\bibinfo {year}
  {2016})}\BibitemShut {NoStop}%
\bibitem [{\citenamefont {Anderson}(1989)}]{ande89}%
  \BibitemOpen
  \bibfield  {author} {\bibinfo {author} {\bibfnamefont {J.~L.}\ \bibnamefont
  {Anderson}},\ }\bibfield  {title} {\enquote {\bibinfo {title} {Colloid
  transport by interfacial forces},}\ }\href {\doibase
  10.1146/annurev.fl.21.010189.000425} {\bibfield  {journal} {\bibinfo
  {journal} {Ann. Rev. Fluid Mech.}\ }\textbf {\bibinfo {volume} {21}},\
  \bibinfo {pages} {61--99} (\bibinfo {year} {1989})}\BibitemShut {NoStop}%
\bibitem [{\citenamefont {Niu}\ \emph {et~al.}(2016)\citenamefont {Niu},
  \citenamefont {Kreissl}, \citenamefont {Brown}, \citenamefont {Rempfer},
  \citenamefont {Botin}, \citenamefont {Holm}, \citenamefont {Palberg},\ and\
  \citenamefont {de~Graaf}}]{niu16}%
  \BibitemOpen
  \bibfield  {author} {\bibinfo {author} {\bibfnamefont {R.}~\bibnamefont
  {Niu}}, \bibinfo {author} {\bibfnamefont {P.}~\bibnamefont {Kreissl}},
  \bibinfo {author} {\bibfnamefont {A.~T.}\ \bibnamefont {Brown}}, \bibinfo
  {author} {\bibfnamefont {G.}~\bibnamefont {Rempfer}}, \bibinfo {author}
  {\bibfnamefont {D.}~\bibnamefont {Botin}}, \bibinfo {author} {\bibfnamefont
  {C.}~\bibnamefont {Holm}}, \bibinfo {author} {\bibfnamefont {T.}~\bibnamefont
  {Palberg}}, \ and\ \bibinfo {author} {\bibfnamefont {J.}~\bibnamefont
  {de~Graaf}},\ }\href@noop {} {\enquote {\bibinfo {title} {Microfluidic
  pumping by micromolar salt concentrations},}\ } (\bibinfo {year} {2016}),\
  \bibinfo {note} {arXiv:1610.00337}\BibitemShut {NoStop}%
\bibitem [{sm()}]{sm}%
  \BibitemOpen
  \href@noop {} {}\bibinfo {note} {See Supplemental Material at xxx for a
  detailed description of the experiments and pH measurements, a calculation of
  the bond vibrations, and more details on the observed structures (citing
  references~\citenum{mali13,perr15}) as well as supporting
  movies.}\BibitemShut {Stop}%
\bibitem [{\citenamefont {Ebbens}\ \emph {et~al.}(2010)\citenamefont {Ebbens},
  \citenamefont {Jones}, \citenamefont {Ryan}, \citenamefont {Golestanian},\
  and\ \citenamefont {Howse}}]{ebbe10}%
  \BibitemOpen
  \bibfield  {author} {\bibinfo {author} {\bibfnamefont {S.}~\bibnamefont
  {Ebbens}}, \bibinfo {author} {\bibfnamefont {R.~A.~L.}\ \bibnamefont
  {Jones}}, \bibinfo {author} {\bibfnamefont {A.~J.}\ \bibnamefont {Ryan}},
  \bibinfo {author} {\bibfnamefont {R.}~\bibnamefont {Golestanian}}, \ and\
  \bibinfo {author} {\bibfnamefont {J.~R.}\ \bibnamefont {Howse}},\ }\bibfield
  {title} {\enquote {\bibinfo {title} {Self-assembled autonomous runners and
  tumblers},}\ }\href {\doibase 10.1103/PhysRevE.82.015304} {\bibfield
  {journal} {\bibinfo  {journal} {Phys. Rev. E}\ }\textbf {\bibinfo {volume}
  {82}},\ \bibinfo {pages} {015304} (\bibinfo {year} {2010})}\BibitemShut
  {NoStop}%
\bibitem [{\citenamefont {Ni}\ \emph {et~al.}(2017)\citenamefont {Ni},
  \citenamefont {Marini}, \citenamefont {Buttinoni}, \citenamefont {Wolf},\
  and\ \citenamefont {Isa}}]{ni17}%
  \BibitemOpen
  \bibfield  {author} {\bibinfo {author} {\bibfnamefont {S.}~\bibnamefont
  {Ni}}, \bibinfo {author} {\bibfnamefont {E.}~\bibnamefont {Marini}}, \bibinfo
  {author} {\bibfnamefont {I.}~\bibnamefont {Buttinoni}}, \bibinfo {author}
  {\bibfnamefont {H.}~\bibnamefont {Wolf}}, \ and\ \bibinfo {author}
  {\bibfnamefont {L.}~\bibnamefont {Isa}},\ }\bibfield  {title} {\enquote
  {\bibinfo {title} {Rational design and fabrication of versatile active
  colloidal molecules},}\ }\href@noop {} {\bibfield  {journal} {\bibinfo
  {journal} {arXiv:1701.08061}\ } (\bibinfo {year} {2017})}\BibitemShut
  {NoStop}%
\bibitem [{\citenamefont {Aldous}(1999)}]{aldo99}%
  \BibitemOpen
  \bibfield  {author} {\bibinfo {author} {\bibfnamefont {David}\ \bibnamefont
  {Aldous}},\ }\bibfield  {title} {\enquote {\bibinfo {title} {Deterministic
  and stochastic models for coalescence (aggregation and coagulation): A review
  of the mean-field theory for probabilists},}\ }\href {\doibase
  10.2307/3318611} {\bibfield  {journal} {\bibinfo  {journal} {Bernoulli}\
  }\textbf {\bibinfo {volume} {5}},\ \bibinfo {pages} {3--48} (\bibinfo {year}
  {1999})}\BibitemShut {NoStop}%
\bibitem [{\citenamefont {Reinm\"uller}\ \emph {et~al.}(2013)\citenamefont
  {Reinm\"uller}, \citenamefont {Sch\"ope},\ and\ \citenamefont
  {Palberg}}]{rein13}%
  \BibitemOpen
  \bibfield  {author} {\bibinfo {author} {\bibfnamefont {A.}~\bibnamefont
  {Reinm\"uller}}, \bibinfo {author} {\bibfnamefont {H.~J.}\ \bibnamefont
  {Sch\"ope}}, \ and\ \bibinfo {author} {\bibfnamefont {T.}~\bibnamefont
  {Palberg}},\ }\bibfield  {title} {\enquote {\bibinfo {title} {Self-organized
  cooperative swimming at low reynolds numbers},}\ }\href {\doibase
  10.1021/la3046466} {\bibfield  {journal} {\bibinfo  {journal} {Langmuir}\
  }\textbf {\bibinfo {volume} {29}},\ \bibinfo {pages} {1738--1742} (\bibinfo
  {year} {2013})}\BibitemShut {NoStop}%
\bibitem [{\citenamefont {Buttinoni}\ \emph {et~al.}(2013)\citenamefont
  {Buttinoni}, \citenamefont {Bialk\'e}, \citenamefont {K\"ummel},
  \citenamefont {L\"owen}, \citenamefont {Bechinger},\ and\ \citenamefont
  {Speck}}]{butt13}%
  \BibitemOpen
  \bibfield  {author} {\bibinfo {author} {\bibfnamefont {I.}~\bibnamefont
  {Buttinoni}}, \bibinfo {author} {\bibfnamefont {J.}~\bibnamefont {Bialk\'e}},
  \bibinfo {author} {\bibfnamefont {F.}~\bibnamefont {K\"ummel}}, \bibinfo
  {author} {\bibfnamefont {H.}~\bibnamefont {L\"owen}}, \bibinfo {author}
  {\bibfnamefont {C.}~\bibnamefont {Bechinger}}, \ and\ \bibinfo {author}
  {\bibfnamefont {T.}~\bibnamefont {Speck}},\ }\bibfield  {title} {\enquote
  {\bibinfo {title} {Dynamical clustering and phase separation in suspensions
  of self-propelled colloidal particles},}\ }\href {\doibase
  10.1103/PhysRevLett.110.238301} {\bibfield  {journal} {\bibinfo  {journal}
  {Phys. Rev. Lett.}\ }\textbf {\bibinfo {volume} {110}},\ \bibinfo {pages}
  {238301} (\bibinfo {year} {2013})}\BibitemShut {NoStop}%
\bibitem [{\citenamefont {Palacci}\ \emph {et~al.}(2013)\citenamefont
  {Palacci}, \citenamefont {Sacanna}, \citenamefont {Steinberg}, \citenamefont
  {Pine},\ and\ \citenamefont {Chaikin}}]{pala13}%
  \BibitemOpen
  \bibfield  {author} {\bibinfo {author} {\bibfnamefont {J.}~\bibnamefont
  {Palacci}}, \bibinfo {author} {\bibfnamefont {S.}~\bibnamefont {Sacanna}},
  \bibinfo {author} {\bibfnamefont {A.~Preska}\ \bibnamefont {Steinberg}},
  \bibinfo {author} {\bibfnamefont {D.~J.}\ \bibnamefont {Pine}}, \ and\
  \bibinfo {author} {\bibfnamefont {P.~M.}\ \bibnamefont {Chaikin}},\
  }\bibfield  {title} {\enquote {\bibinfo {title} {Living crystals of
  light-activated colloidal surfers},}\ }\href {\doibase
  10.1126/science.1230020} {\bibfield  {journal} {\bibinfo  {journal}
  {Science}\ }\textbf {\bibinfo {volume} {339}},\ \bibinfo {pages} {936--940}
  (\bibinfo {year} {2013})}\BibitemShut {NoStop}%
\bibitem [{\citenamefont {Yan}\ \emph {et~al.}(2016)\citenamefont {Yan},
  \citenamefont {Han}, \citenamefont {Zhang}, \citenamefont {Xu}, \citenamefont
  {Luijten},\ and\ \citenamefont {Granick}}]{yan16}%
  \BibitemOpen
  \bibfield  {author} {\bibinfo {author} {\bibfnamefont {J.}~\bibnamefont
  {Yan}}, \bibinfo {author} {\bibfnamefont {M.}~\bibnamefont {Han}}, \bibinfo
  {author} {\bibfnamefont {J.}~\bibnamefont {Zhang}}, \bibinfo {author}
  {\bibfnamefont {C.}~\bibnamefont {Xu}}, \bibinfo {author} {\bibfnamefont
  {E.}~\bibnamefont {Luijten}}, \ and\ \bibinfo {author} {\bibfnamefont
  {S.}~\bibnamefont {Granick}},\ }\bibfield  {title} {\enquote {\bibinfo
  {title} {Reconfiguring active particles by electrostatic imbalance},}\ }\href
  {\doibase 10.1038/nmat4696} {\bibfield  {journal} {\bibinfo  {journal} {Nat.
  Mater.}\ }\textbf {\bibinfo {volume} {advance online publication}} (\bibinfo
  {year} {2016}),\ 10.1038/nmat4696}\BibitemShut {NoStop}%
\bibitem [{\citenamefont {Zhang}\ \emph {et~al.}(2016)\citenamefont {Zhang},
  \citenamefont {Yan},\ and\ \citenamefont {Granick}}]{zhan16}%
  \BibitemOpen
  \bibfield  {author} {\bibinfo {author} {\bibfnamefont {J.}~\bibnamefont
  {Zhang}}, \bibinfo {author} {\bibfnamefont {J.}~\bibnamefont {Yan}}, \ and\
  \bibinfo {author} {\bibfnamefont {S.}~\bibnamefont {Granick}},\ }\bibfield
  {title} {\enquote {\bibinfo {title} {Directed self-assembly pathways of
  active colloidal clusters},}\ }\href {\doibase 10.1002/anie.201509978}
  {\bibfield  {journal} {\bibinfo  {journal} {Angew. Chem. Int. Ed.}\ }\textbf
  {\bibinfo {volume} {55}},\ \bibinfo {pages} {5166--5169} (\bibinfo {year}
  {2016})}\BibitemShut {NoStop}%
\bibitem [{\citenamefont {Sundararajan}\ \emph {et~al.}(2008)\citenamefont
  {Sundararajan}, \citenamefont {Lammert}, \citenamefont {Zudans},
  \citenamefont {Crespi},\ and\ \citenamefont {Sen}}]{sund08}%
  \BibitemOpen
  \bibfield  {author} {\bibinfo {author} {\bibfnamefont {S.}~\bibnamefont
  {Sundararajan}}, \bibinfo {author} {\bibfnamefont {P.~E.}\ \bibnamefont
  {Lammert}}, \bibinfo {author} {\bibfnamefont {A.~W.}\ \bibnamefont {Zudans}},
  \bibinfo {author} {\bibfnamefont {V.~H.}\ \bibnamefont {Crespi}}, \ and\
  \bibinfo {author} {\bibfnamefont {A.}~\bibnamefont {Sen}},\ }\bibfield
  {title} {\enquote {\bibinfo {title} {Catalytic motors for transport of
  colloidal cargo},}\ }\href {\doibase 10.1021/nl072275j} {\bibfield  {journal}
  {\bibinfo  {journal} {Nano Letters}\ }\textbf {\bibinfo {volume} {8}},\
  \bibinfo {pages} {1271--1276} (\bibinfo {year} {2008})}\BibitemShut {NoStop}%
\bibitem [{\citenamefont {Baraban}\ \emph {et~al.}(2012)\citenamefont
  {Baraban}, \citenamefont {Tasinkevych}, \citenamefont {Popescu},
  \citenamefont {Sanchez}, \citenamefont {Dietrich},\ and\ \citenamefont
  {Schmidt}}]{bara12}%
  \BibitemOpen
  \bibfield  {author} {\bibinfo {author} {\bibfnamefont {L.}~\bibnamefont
  {Baraban}}, \bibinfo {author} {\bibfnamefont {M.}~\bibnamefont
  {Tasinkevych}}, \bibinfo {author} {\bibfnamefont {M.~N.}\ \bibnamefont
  {Popescu}}, \bibinfo {author} {\bibfnamefont {S.}~\bibnamefont {Sanchez}},
  \bibinfo {author} {\bibfnamefont {S.}~\bibnamefont {Dietrich}}, \ and\
  \bibinfo {author} {\bibfnamefont {O.~G.}\ \bibnamefont {Schmidt}},\
  }\bibfield  {title} {\enquote {\bibinfo {title} {Transport of cargo by
  catalytic janus micro-motors},}\ }\href {\doibase 10.1039/C1SM06512B}
  {\bibfield  {journal} {\bibinfo  {journal} {Soft Matter}\ }\textbf {\bibinfo
  {volume} {8}},\ \bibinfo {pages} {48--52} (\bibinfo {year}
  {2012})}\BibitemShut {NoStop}%
\bibitem [{\citenamefont {Malins}\ \emph {et~al.}(2013)\citenamefont {Malins},
  \citenamefont {Williams}, \citenamefont {Eggers},\ and\ \citenamefont
  {Royall}}]{mali13}%
  \BibitemOpen
  \bibfield  {author} {\bibinfo {author} {\bibfnamefont {A.}~\bibnamefont
  {Malins}}, \bibinfo {author} {\bibfnamefont {S.~R.}\ \bibnamefont
  {Williams}}, \bibinfo {author} {\bibfnamefont {J.}~\bibnamefont {Eggers}}, \
  and\ \bibinfo {author} {\bibfnamefont {C.~P.}\ \bibnamefont {Royall}},\
  }\bibfield  {title} {\enquote {\bibinfo {title} {Identification of structure
  in condensed matter with the topological cluster classification},}\ }\href
  {\doibase http://dx.doi.org/10.1063/1.4832897} {\bibfield  {journal}
  {\bibinfo  {journal} {J. Chem. Phys.}\ }\textbf {\bibinfo {volume} {139}},\
  \bibinfo {pages} {234506} (\bibinfo {year} {2013})}\BibitemShut {NoStop}%
\end{thebibliography}
\end{document}